\documentclass[12pt]{article}
\usepackage{amssymb,amsmath}
\usepackage[noblocks]{authblk}
\usepackage[top=0.75in, bottom=0.75in, left=0.75in, right=0.75in, dvips]{geometry}
\usepackage{caption}  
\date{}   

\begin{document}
\textwidth 10.0in       
\textheight 9.0in 
\topmargin -0.60in
\title{Renormalization Scheme and Mass Scale Independence}
\author[1]{F.A. Chishtie \thanks{Email: fachisht@uwo.ca}}
\author[1,2] {D.G.C. McKeon \thanks{Email: dgmckeo2@uwo.ca}}
\affil[1] {Department of Applied Mathematics, The
University of Western Ontario, London, ON N6A 5B7, Canada}
\affil[2] {Department of Mathematics and
Computer Science, Algoma University, Sault Ste. Marie, ON P6A
2G4, Canada}  
\maketitle    

\maketitle
\noindent

\begin{abstract}
We demonstrate that in the mass independent renormalization scheme, the renormalization group equations associated with the unphysical parameters that characterize the renormalization scheme and the mass scale leads to summation that results in a cancellation between the implicit and explicit dependence on these parameters.  The resulting perturbative expansion is consequently independent of these arbitrary parameters.  We illustrate this by considering $R$, the cross section for $e^+e^- \rightarrow$ hadrons.
\end{abstract}

\section{Introduction}

When a perturbative calculation of the radiative effects contributing to $R$, the cross section for $e^+e^- \rightarrow$ hadrons, is carried out to some finite order, then the result is dependent on the renormalization scheme used, even within mass independent renormalization schemes [1,2].  These ambiguities are characterized in several ways.  A mass scale $\mu$ naturally arises when one renormalizes when using dimensional regularization [3,4,5].  The renormalization, when using dimensional regularization, involves absorbing poles in $\epsilon = 2 - n/2$ ($n$ - number of dimensions) that arise in perturbation theory into the coupling $a$, if we restrict our attention to massless QCD. Consequently, $a$ then acquires implicit dependence on $\mu$.  In addition to the ambiguity residing in this unphysical mass scale $\mu$, there is an ambiguity occurring on account of the possibility of performing a finite renormalization in which $a$ is replaced by $a^{\prime}$ where
\begin{equation}
a^\prime = a + x_2 a^2 + x_3 a^3 + \ldots .
\end{equation}
This ambiguity can either be characterized by the parameters $x_3, x_4 \ldots$, or the coefficients $c_2, c_3 \ldots$ of the function
\begin{equation}
\beta(a) = -ba^2 \left( 1+ ca + c_2 a^2 + \ldots \right)
\end{equation}
that governs how $a$ varies with $\mu$ [6]
\begin{equation}
\mu \frac{\partial a}{\partial\mu} = \beta(a).
\end{equation}
Any physical quantity $R$, if it were known exactly, must be independent of such unphysical parameters; this leads to the renormalization group equations
\begin{equation}\tag{4a,b}
\mu \frac{dR}{d\mu} = \frac{dR}{dc_i} = 0.
\end{equation}
If a perturbative calculation of $R$ is carried out to N$^{th}$ order in $a$ in perturbation theory so that
\begin{equation}\tag{5}
R^{(N)} = \sum_{n=0}^{N} \rho_n a^{n+1}
\end{equation}
then $R^{N}$ has explicit dependence on $(\mu, c_i)$ through $\rho_n$ and implicit dependence on these parameters through $a$.  Quite often some choice is made of these parameters (eg, working with $\overline{MS}$ [7] renormalization with $\mu$ chosen to be the centre of mass energy $Q$).  We shall show how eq. (4a) can be used to sum the explicit dependence of $R$ on $\mu$ to all orders in perturbation theory, and how this leads to a cancellation between the implicit and explicit dependence of $R$ on $\mu$ [8,9, 10].  Next we shall follow ref. [11, 12] and show how renormalization scheme dependence on the parameters $c_i$ can be replaced by dependence on a single parameter $\gamma$ once the expansion parameter $a$ is replaced by a parameter $z$.  Once this is done, we will show how the condition
\begin{equation}\tag{6}
\frac{dR}{d\gamma} = 0
\end{equation}
can be used to sum the explicit dependence of $R$ on $\gamma$; as with $\mu$, the explicit and implicit dependence of $R$ on $\gamma$ is found to then cancel.

\section{Dependence on Mass Scale}

When using dimensional regularization and a mass independent renormalization scheme, the coupling $a$ has implicit dependence on an unphysical mass parameter $\mu$ so that we have eqs. (2,3) satisfied.  Following ref. [6], we write the solution to this equation as
\begin{equation}\tag{7}
\ln \frac{\mu}{\Lambda} = \int_0^a \frac{dx}{\beta(x)} + \int_0^\infty \frac{dx}{bx^2(1+cx)}
\end{equation}
where $\Lambda$ is a mass parameter introduced as part of the boundary condition on eq. (3).  The second integral in eq. (7) serves to cancel divergences at $x = 0$ in the first integral.

From eq. (7) we see that $a$ in eq. (5) has implicit dependence on $\mu$.  Furthermore, the nature of perturbative calculations shows that $\rho_n$ in eq. (5) has the form
\begin{equation}\tag{8}
\rho_n = \sum_{k=0}^n  T_{n,k} \left( \ln \frac{\mu}{Q}\right)^k.
\end{equation}
From eqs. (5) and (8), we can write
\begin{equation}\tag{9}
R = \sum_{m=0}^\infty A_m(a)\left( \ln \frac{\mu}{Q}\right)^m
\end{equation}
where
\begin{equation}\tag{10}
A_m (a) = \sum_{k=0}^\infty T_{m+k, m} a^{m +k+1} .
\end{equation}
We note that presenting the dependence of $R$ on $\mu$ in this way is distinct from using the ``method of characteristics'' approach first introduced in [13](see also [14]); rather it relies on the form of results that follow from applying the renormalization procedure to eliminate divergences arising in Feynman integrals.

We now can write eq. (4a) as
\begin{equation}\tag{11}
\left( \mu \frac{\partial}{\partial\mu} + 
\beta(a) \frac{\partial}{\partial a}\right) \sum_{m=0}^\infty A_m(a) \left( \ln \frac{\mu}{Q}\right)^m = 0
\end{equation}
which shows that
\begin{equation}\tag{12}
A_m\left(a \ln \left(\frac{\mu}{\Lambda}\right)\right) = - \frac{\beta\left(a \left(\ln\frac{\mu}{\Lambda}\right)\right)}{m} \frac{d}{da}
A_{m-1} \left(a  \left( \ln\frac{\mu}{\Lambda}\right)\right) 
\end{equation}
which by eq. (3) becomes
\begin{equation}\tag{13}
= - \frac{1}{m} \;\frac{d}{d\left(\ln \frac{\mu}{\Lambda}\right)} A_{mn} \left(a \left(\ln \frac{\mu}{\Lambda}\right)\right).
\end{equation}
This can be iterated so that
\begin{equation}\tag{14}
A_m(a) = \frac{(-1)^m}{m!}\frac{d^m}{\left( d\ln \frac{\mu}{\Lambda}\right)^m} A_0(a)
\end{equation}
and so eq. (9) becomes
\begin{equation}
R = \sum_{m=0}^\infty \frac{1}{m!}\left( -\ln \frac{\mu}{Q}\right)^m \frac{d^m}{d\left(\ln \frac{\mu}{\Lambda}\right)^m} A_0
\left( a \left( \ln\frac{\mu}{\Lambda}\right)\right)\nonumber
\end{equation}
\begin{equation}\tag{15}
\hspace{-3cm} = A_0 \left( a \left( \ln\frac{\mu}{\Lambda} - \ln \frac{\mu}{Q}\right)\right).
\end{equation}
As $- \ln\frac{\mu}{Q} + \ln \frac{\mu}{\Lambda} = \ln \frac{Q}{\Lambda}$, the implicit and explicit dependence of $R$ on the unphysical mass scale parameter $\mu$ has cancelled in eq. (15).  $R$ is given in terms of the log independent contributions at each order in powers of $a$ with $a$ evaluated at $\ln\frac{Q}{\Lambda}$ so that
\begin{equation}\tag{16}
R = \sum_{n=0}^\infty T_n a\left( \ln \frac{Q}{\Lambda}\right)^{n+1} \quad (T_n \equiv T_{n,0} ).
\end{equation}

Having demonstrated how the renormalization group equation can be used to eliminate dependence of $R$ on $\mu$, we will now show how a similar argument can be employed to eliminate dependence of $R$ on the renormalization scheme being used.

\section{Dependence on Renormalization Scheme}

If under the finite renormalization of eq. (1), $a^\prime$ satisfies
\begin{equation}\tag{17}
\mu \frac{\partial a^\prime}{\partial\mu} = - b^\prime a^{\prime^{2}} \left[ 1 + c^\prime a^\prime + c_2^\prime a^{\prime^{2}} + c_3^\prime a^{\prime^{3}}
 + \ldots \right]
\end{equation}
then together eqs. (1, 2, 3, 17) show that [15]
\begin{equation}\tag{18a}
b^\prime = b
\end{equation}
\begin{equation}\tag{18b}
c^\prime = c
\end{equation}
\begin{equation}\tag{18c}
c_2^\prime = c_2 - cx_2 + x_3 - x_2^2
\end{equation}
\begin{equation}\tag{18d}
c_3^\prime = c_3 - 3cx_2^2 + 2 \left( c_2 - 2c_2^\prime\right) x_2 + 2x_4 - 2x_2x_3
\end{equation}
\begin{equation}\tag{18e}
c_4^\prime = c_4 - 2x_4x_2 - x_3^2 + c \left( x_4 - x_2^3 - 6x_2x_3\right) + 3x_3c_2 - 4x_3 c_2^\prime
\end{equation}
\begin{equation}
-6x_2^2 c_2^\prime + 2x_2c_3 - 5x_2 c_3^\prime + 3x_5\nonumber
\end{equation}
etc.\\
Eq. (18) shows that $x_n(n \geq 3)$ can be expressed in terms of $c_2, c_2^\prime, c_3, c_3^\prime , \ldots c_{n-1}, c_{n-1}^\prime$ and
$x_2$ but so that the renormalization scheme ambiguity of eq. (1) can be characterized by the coefficients $c_2, c_3 \ldots$ of $\beta(a)$ in eq. (2), as well as the coefficient $x_2$ of eq. (1) [6].  This parameter $x_2$ is related to the mass scale $\mu$.  To see this, we first make the expansion
\begin{equation}\tag{19}
a(L^\prime) = a(L) \Big[ 1 + \sigma_{11} \lambda a(L) + \left( \sigma_{21} \lambda + \sigma_{22} \lambda^2\right)a^2(L)
\end{equation}
\begin{equation}
+ \left(\sigma_{31} \lambda + \sigma_{32} \lambda^2 + \sigma_{33} \lambda^3\right) a^3 (L) + \ldots \Big]\nonumber
\end{equation}
where $L = \ln \frac{\mu}{\Lambda}$,  $L^\prime = \ln \frac{\mu^\prime}{\Lambda}$ and  $\lambda = \ln \frac{\mu}{\mu^\prime}$.  Since
$\mu\frac{d}{d\mu} a(L^\prime) = 0$, eqs. (2, 3, 19) together lead to [16]
\begin{equation}\tag{20}
a(L^\prime) = a(L) \left[ 1 + b\lambda \left( a(L) + (c+b\lambda) a^2(L) + \left(c_2 + \frac{5}{2} bc\lambda + b^2 \lambda^2\right) a^3(L) + \ldots \right) \right] .
\end{equation}
If now in eq. (18) we were to set $c_n = c_n^\prime (n = 2,3,\ldots)$ and then put the solutions for $x_3, x_4 \ldots$ in terms of $c_2, c_3 \ldots , x_2$ into eq. (1), we recover eq. (20) provided $x_2$ is identified with $b \ln \frac{\mu}{\mu^\prime}$.  This shows how the variation of $x_2$ is related to the variation of $\mu$.

This can be illustrated in another way.  We begin by slightly altering eq. (7) so that now
\begin{equation}\tag{21}
L \equiv \ln \frac{\mu}{\Lambda} = \frac{1}{b}\left[ \frac{1}{a} - c \ln \left( \frac{1+ca}{a}\right)\right] + \int_0^a dx \left( \frac{1}{\beta(x)} + \frac{1}{bx^2(1+cx)}\right).
\end{equation}
If now in eq. (1), $a$ and $a^\prime$ are evaluated at the same value of $\mu$ and different values of $\Lambda$ (or, alternatively, different values of $\mu$ and $\mu^\prime$ and the same value of $\Lambda$ and $\Lambda^\prime$) then by eq. (21) [6, 17]
\begin{equation}\tag{22}
L^\prime - L = \frac{-x_2}{b} + \mathcal{O}(a).
\end{equation}
Since eq. (22) is true for all $a$, it must be true in the limit $a \rightarrow 0$, which is consistent with identifying $x_2$ with $b \ln \frac{\mu}{\mu^\prime}$ in eq. (20).  This leads to $\Lambda / \Lambda^\prime = e^{-x_2/b}$ [17].

In ref. [11,12], a slightly different definition of $L$ was used,
\begin{equation}\tag{23}
\overline{L} = \ln \frac{\mu}{\overline{\Lambda}} = \frac{1}{b} \left[ \frac{1}{a} + c \ln a\right] + \int_0^a dx \left( \frac{1}{\beta(x)}
+ \frac{1}{bx^2} (1-cx)\right).
\end{equation}

We now consider how $a$ varies when $c_n(n = 2,3, \ldots)$ is altered.  Since
\begin{equation}\tag{24}
\frac{\partial^2a}{\partial\mu \partial c_i} = \frac{\partial^2 a}{\partial c_i \partial_\mu},
\end{equation}
eq. (3) leads to [6]
\begin{equation}\tag{25}
\frac{\partial a}{\partial c_i} = B_i(a, c_1)
\end{equation}
with
\begin{equation}\tag{26}
B_i (a, c_i) = - b \beta(a) \int_0^a dx \frac{x^{i+2}}{\beta^2(x)}.
\end{equation}
We see from eq. (26) that from eqs. (21) and (23) that
\begin{equation}\tag{27}
\frac{\partial L}{\partial c_i} = \frac{\partial \overline{L}}{\partial c_i} = 0
\end{equation}
showing that $\mu/\Lambda$ and $\mu/\overline{\Lambda}$ are both independent of the $c_i$ and hence of $x_i( i \geq 3)$; they are altered only if $x_2$ is changed in eq. (1).

Following ref. [11, 12, 18] we use this result to define a new coupling $z$ to replace $a$ so that all renormalization scheme dependence  of $z$ resides in a single parameter $\gamma$.  Working from eq. (21) (rather than eq. (23) as with refs. [11, 12, 18]) we set
\begin{equation}\tag{28}
\frac{1}{b} \left[ \frac{1}{z} -c\ln \left(\frac{1+cz}{z}\right)\right] - \ln \gamma = 
\frac{1}{b} \left[ \frac{1}{a} -c\ln \left(\frac{1+ca}{a}\right)\right] + \int_0^a dx
\left( \frac{1}{\beta(x)} + \frac{1}{bx^2(1+cx)}\right).
\end{equation}
With $z = z\left( \ln\frac{\mu}{\Lambda}, \ln \gamma\right)$ it now follows from eq. (28) that
\begin{equation}\tag{29a}
\mu \frac{\partial z}{\partial \mu} \left( \ln \frac{\mu}{\Lambda}, \ln \gamma \right) = \beta_0 \left(z \left( \ln \frac{\mu}{\Lambda}, \ln \gamma \right)\right)
\end{equation}
and
\begin{equation}\tag{29b}
\gamma \frac{\partial z}{\partial \gamma} \left( \ln \frac{\mu}{\Lambda}, \ln \gamma \right) = \beta_0 \left(z \left( \ln \frac{\mu}{\Lambda}, \ln \gamma \right)\right)
\end{equation}
where
\begin{equation}\tag{30}
\beta_0 (z) = - b z^2(1+cz).
\end{equation}
(An analogous calculation following from eq. (23) gives
\begin{equation}\tag{31}
\frac{1}{b} \left[ \frac{1}{\overline{z}} + c \ln \overline{z}\right] - \overline{\gamma} = \frac{1}{b} \left[ \frac{1}{a} + 
c \ln a\right] + \int_0^a dx \left( \frac{1}{\beta(x)}+ \frac{1}{bx^2} (1-cx)\right)
\end{equation}
so that
\begin{equation}\tag{32a,b}
\mu \frac{\partial\overline{z}}{\partial \mu} = \overline{\beta}_0 (\overline{z}) = \frac{\partial\overline{z}}{\partial \overline{\gamma}}
\end{equation}
where
\begin{equation}\tag{33}
\overline{\beta}_0 (z) = \frac{-bz^2}{1-cz}
\end{equation}
as in ref. [11, 12, 18].) \\
The function $\beta_0$ occurs when one uses 't Hooft renormalization [19] in which $c_i = 0 (i \geq 2)$ while $\overline{\beta}_0$ occurs in some ways of renormalizing $N = 1$ super Yang-Mills theory [20, 21].

From eq. (26), it follows that [22]
\begin{equation}\tag{34}
a(c_i^\prime) = a(c_i) + (c_2^\prime - c_2) a^2 (c_i) + \frac{1}{2} (c_3^\prime - c_3) a^3 (c_i) + \Big[ \frac{1}{6} (c_2^{\prime^2} - c_2^2)
\end{equation}
\begin{equation}
+\frac{3}{2} (c_2^\prime - c_2) - \frac{c}{6}(c_3^\prime - c_3) + \frac{1}{3} (c_4^\prime - c_4)\Big] a^4 (c_i) + \ldots \nonumber
\end{equation}
This is a finite renormalization that is distinct from that of eq. (1); it comes from the requirement that
\begin{equation}\tag{35}
\frac{d}{dc_i} a(c_i^\prime) = \left( \frac{\partial}{\partial c_i} + B_i (a, c_i) \frac{\partial}{\partial a}\right) a(c_i^\prime) = 0.
\end{equation}
If now in eq. (16) we have the condition
\begin{equation}\tag{36}
\frac{dR}{dc_i} = 0 = \left( \frac{\partial}{\partial c_i} + B_i (a, c_i) \frac{\partial}{\partial a}\right) \sum_{n=0}^\infty T_n (c_i) a^{n+1} (c_i)
\end{equation}
then we see that [22]
\begin{equation}\tag{37a-f}
\hspace{-1cm}T_0 = \tau_0, \quad  T_1 = \tau_1,\quad T_2 = -c_2 + \tau_2, \quad T_3 = -2 c_2\tau_1-\frac{1}{2} c_3 + \tau_3
\end{equation}
\begin{equation}
\hspace{-3.5cm}T_4 = -\frac{1}{3} c_4 - \frac{c_3}{2} \left(- \frac{1}{3} c+ 2\tau_1\right) + \frac{4}{3} c_2^2- 3 c_2\tau_2 + \tau_4\nonumber
\end{equation}
\begin{equation}
T_5 = \left[ \frac{1}{3} cc_2^2 + \frac{3}{2}c_2c_3 + \frac{11}{3} c_2^2 \tau_1 - 4c_2\tau_3 \right] - \frac{1}{2} 
\left[ \frac{1}{6} c^2c_3 - \frac{2}{3} c_3 c \tau_1 + 3c_3\tau_2\right]\nonumber
\end{equation}
\begin{equation}
-\frac{1}{3} \left[ -\frac{1}{2} c_4c + \frac{1}{2} c_4 \tau_1\right] - \frac{1}{4} c_5 + \tau_5 \nonumber
\end{equation}
etc. \\
where $\tau_0, \tau_1 \ldots$ are constants that are renormalization scheme invariant. Having determined by eq. (37) the dependence of each of the $T_n$ on $c_i$, one would expect that by somehow doing the sum in eq. (16) then the explicit dependence of $R$ on $c_i$ through eq. (37) would cancel against the implicit dependence of $R$ on $c_i$ through $a(c_i)$ as formed by eqs. (25) and (26).  Unfortunately, it is not apparent how this summation could be done as there is no analogue of eq. (8).  This is because the dependence of $T_n$ on $c_n$ is highly non-linear, while in eq. (8) we have $\rho_\mu$ depending on $\ln \frac{\mu}{\Lambda}$ through a simple polynomial.

However, we have seen that $z$ introduced in eq. (28) has all renormalization scheme dependence residing in the single parameter $\ln \gamma$.  Furthermore, since $z$ satisfies eq. (29a), we can use eq. (34) to find $a$ in terms of $z$ by taking $c_i = 0$,
\begin{equation}\tag{38}
a(c_i) = z(\ln \gamma) + c_2 z(\ln \gamma)^2 + \frac{1}{2} c_3^2 z(\ln \gamma)^3 + 
\left[ \frac{1}{6} c_2^2 + \frac{3}{2} c_2 - \frac{c}{6} c_3 + \frac{1}{3} c_4\right] z(\ln \gamma)^4 + \ldots
\end{equation}
Substitution of eq. (37) into eq. (16) leads to 
\begin{equation}
R = \sum_{n=0}^\infty T_n \left( z(\ln \gamma) + c_2 z(\ln \gamma)^2 + \frac{1}{2} c_3^2 z(\ln\gamma)^3 + \ldots \right)^{n+1} \nonumber
\end{equation}
\begin{equation}\tag{39}
\hspace{-4.3cm}= \sum_{m+0}^\infty U_n (\ln \gamma) z(\ln\gamma)^{n+1} .
\end{equation}
However, now we have $z$ having both mass scale dependence and renormalization scheme dependence governed by the same function $\beta_0$ according to eqs. (29, 30) and so from eq. (21) we have
\begin{equation}\tag{40}
\ln \left(\frac{\mu}{\Lambda}\right) + \ln \gamma = \ln  \left(\frac{\mu\gamma}{\Lambda}\right) = \frac{1}{b} 
\left[ \frac{1}{z} - c\ln \left(\frac{1+cz}{z}\right)\right].
\end{equation}
(This results in $z$ being given by a Lambert $W$ function.)  We thus have $\mu$ and $\gamma$ only occurring in the product $\mu\gamma$; from this we infer that summing the explicit dependence of $R$ on $\mu$ to cancel against its implicit dependence of $R$ on $\mu$ through $z(\mu)$ will also lead to a cancellation of the implicit and explicit dependence of $R$ on $\gamma$ when using $z$ as an expansion parameter.  From eq. (16) we then have
\begin{equation}\tag{41}
R = \sum_{n=0}^\infty T_n^{(0)} z \left( \ln \frac{Q}{\Lambda}\right)^{n+1}.
\end{equation}
Since $z$ satisfies eq. (29), in which $c_i = 0 (i \geq 2)$ by eq. (30), we see by eq. (37) that in eq. (41)
\begin{equation}\tag{42}
T_n^{(0)} = \tau_n
\end{equation}
which are invariant under a change of renormalization scheme.  In eq. (41) we have an expression for $R$ that is independent of both the unphysical mass scale $\mu$ introduced in the process of renormalization and free of any ambiguities arising from making a finite renormalization.

One might anticipate the result of eq. (41), as if one were able to use eq. (4b) to sum all dependence of $R$ on $c_i$, then the cancellation that should result between the implicit and explicit dependence of $R$ on $c_i$ would hold if $c_i = 0$.  But if $c_i = 0$, then eq. (16) reduces to eq. (41).

\section{Discussion}
Perturbative calculations when carried out to finite order result in expressions for physical quantities such a $R$ having dependence on an unphysical mass scale and the choice of renormalization scheme.  The way in which Feynman diagrams are computed makes it possible to know the way in which each order in perturbation theory depends on the unphysical mass scale; one can then use the renormalization group equation associated with this mass scale to sum to all orders this explicit dependence on mass scale.  We then find that this explicit dependence cancels against the implicit dependence which resides in the coupling, which is the parameter used in the perturbative expansion.

It is not immediately possible to apply this procedure to eliminate dependence of any perturbative calculation on those parameters $c_i$ that characterize a renormalization scheme.  This is because at each order of perturbation theory the dependence on these parameters does not lend itself to summation by using the associated renormalization group equation.  However, in refs. [11, 12] it is shown that one can replace the coupling $a$ with another expansion parameter $z$ so that the renormalization scheme is now characterized by a single parameter $\gamma$ and that the renormalization group equation associated with this parameter makes it possible to show that, just as with the unphysical mass scale, the implicit and explicit dependence of $R$ on it cancels. Recently, an alternative to eliminating renormalization scale and scheme dependence in observables based on Effective Field Theory techniques using a newly devised ``Principle of Observable Effective Matching" (POEM) [25]. In POEM, the focus is directly on physical observables, and expressions from ambiguities are derived via matching of the scale and scheme dependent expressions at a relevant physical scale. As such, ``Effective Physical Observables" (EPO) are derived, which are at a known loop order in perturbation theory. We will explore the possible connections of this work with the POEM aproach in upcoming work.  

The final result for $R$ in eq. (41) is independent of all unphysical parameters induced by renormalization.  The coupling $z$ is dependent on $b$ and $c$, both of which have been long known.  To obtain $T_n^{(0)} = \tau_n$, we need to compute $T_n$ and $c_2 \ldots c_n$ in eq. (37) which are computed in some convenient renormalization scheme such as $\overline{MS}$.

We wish to extend this approach to scheme dependence to deal with situations involving multiple couplings [23] or to problems such as the QCD static potential [26].

\section*{Acknowledgements}
Correspondence with A. Kataev has been very stimulating.  R. Macleod made a useful suggestion.

\section*{Appendix-Mass Parameters}

In ref. [10], the renormalization mass scale and renormalization scheme dependency of a perturbative calculation of a physical quantity was considered when a massive particle is involved in the process being considered (which is in this case the inclusive semi-leptonic $b$ decays).  We wish to briefly address in this appendix how the approach outlined in this paper can be applied when a massive particle occurs and one uses mass independent renormalization.

If a mass $m$ is renormalized using a mass independent renormalization scheme, then
\begin{equation}
\hspace{-6.4cm}\mu \frac{dm(\mu)}{d\mu} = m(\mu) \delta(a(\mu))\nonumber
\end{equation}
\begin{equation}\tag{A.1}
= m(\mu) f a(\mu) \left( 1 + g_1 a(\mu) + g_2a^2(\mu) + \ldots\right)
\end{equation}
where $f$ is renormalization scheme invariant under the change of eq. (1) along with the finite renormalization of $m$
\begin{equation}\tag{A.2}
m^\prime = m \left(1 + y_1 a + y_2 a^2 + \ldots\right).
\end{equation}
However, the coefficients $g_1, g_2 \ldots$ in eq. (A.1) can be seen to be scheme dependent, and much like the coefficients $c_2, c_3 \ldots$ in eq. (2).  We find that along with eq. (25) we have [24]
\begin{equation}\tag{A.3}
\frac{\partial m}{\partial a_i}= m \left(\frac{\delta(a)}{\beta(a)} B_i(a) + b \int_0^a dx \frac{x^{i+2}\delta(x)}{\beta^2(x)}\right)
\end{equation}
and 
\begin{equation}\tag{A.4}
\frac{\partial m}{\partial g_i} = m f \int_0^a dx \frac{x^{i+1}}{\beta(x)}.
\end{equation}
Together, eqs. (3) and (A.1) show that
\begin{equation}\tag{A.5}
\frac{dm}{da} = \frac{m\delta(a)}{\beta(a)}
\end{equation}
which,in analogy with eq. (7) leads to 
\begin{equation}\tag{A.6}
m(a(\mu)) = I\!\!M \exp \left[ \int_0^a dx \left( \frac{\delta (x)}{\beta(x)}\right) + \int_0^\infty dx \left( \frac{fx}{bx^2(1+cx)}\right)\right].
\end{equation}
Together, eqs. (25, A.3, A.4) show that the mass parameter $I\!\!M$ in eq. (26) is independent of $\mu$, $g_i$, $c_i$, much like $L$ in eq. (21) is independent of $c_i$.

Just as eq. (7) is replaced by eq. (21), we replace eq. (A.6) by
\begin{equation}\tag{A.7}
m(a(\mu)) = I\!\!M \exp \left[ \frac{f}{b}\ln \left( \frac{1+ca}{a}\right) + \int_0^a dx \left( \frac{\delta (x)}{\beta(x)} + \frac{fx}{bx^2(1+cx)}\right)\right].
\end{equation}
We now introduce a mass function $\kappa(z)$ to replace $m(a)$ with $z$ defined by eq. (28),
\begin{equation}\tag{A.8}
\frac{\kappa(z)}{\Phi \exp \left( \frac{f}{b} \ln \left( \frac{1+cz}{z}\right)\right)} = 
\frac{m(a)}{I\!\!M \exp \left[ \frac{f}{b}\ln \left( \frac{1+ca}{a}\right) + \int_0^a dx \left( \frac{\delta(x)}{\beta(x)} + \frac{fx}{bx^2(1+cx)}\right)\right]}
\end{equation}
where $\Phi$ is a massive parameter much like $\gamma$ in eq. (28).  From eq. (A.8) we see that
\begin{equation}\tag{A.9a,b}
\frac{\partial \kappa}{\partial c_i} = \frac{\partial \kappa}{\partial g_i} = 0
\end{equation}
while
\begin{equation}\tag{A.10}
\mu \frac{\partial \kappa}{\partial \mu}  = fz = \gamma \frac{\partial \kappa}{\partial \gamma}
\end{equation}
\begin{equation}\tag{A.11}
\Phi \frac{\partial \kappa}{\partial \Phi} = \kappa .
\end{equation}
As a result, we see that all renormalization scheme dependence of $\kappa$ now resides in the parameters $\gamma$ and $\Phi$.

\end{document}